\def\year{2022}\relax
\documentclass[letterpaper]{article} % DO NOT CHANGE THIS
\usepackage{aaai22}  % DO NOT CHANGE THIS
\usepackage{times}  % DO NOT CHANGE THIS
\usepackage{helvet}  % DO NOT CHANGE THIS
\usepackage{courier}  % DO NOT CHANGE THIS
\usepackage[hyphens]{url}  % DO NOT CHANGE THIS
\usepackage{graphicx} % DO NOT CHANGE THIS
\usepackage{natbib}  % DO NOT CHANGE THIS AND DO NOT ADD ANY OPTIONS TO IT
\usepackage{caption} % DO NOT CHANGE THIS AND DO NOT ADD ANY OPTIONS TO IT
\DeclareCaptionStyle{ruled}{labelfont=normalfont,labelsep=colon,strut=off} % DO NOT CHANGE THIS
\usepackage{algorithm}
\usepackage{newfloat}
\usepackage{listings}
\floatstyle{ruled}
\newfloat{listing}{tb}{lst}{}
\floatname{listing}{Listing}

\usepackage[utf8]{inputenc}

\usepackage{url}
\usepackage{appendix}
\usepackage{graphicx}
\usepackage{pgfplots}
\usepackage{subcaption}
\usepackage{tkz-euclide}
\usepackage{bbm}
\usepackage{multirow}
\usepackage{amsmath}
\usepackage{amsfonts}
\usepackage{dsfont}
\usetikzlibrary{positioning}
\usepackage{amsthm}
\usepackage[noend]{algpseudocode}

\usepackage{tikz}
\usepackage{xcolor}
\usepackage{tkz-euclide}
\usetikzlibrary{shapes}
\usepackage{pgfplots}
\pgfplotsset{compat=newest}
\usetikzlibrary{shapes.arrows}
\usetikzlibrary{trees,arrows}
\usetikzlibrary{arrows.meta,
                positioning,
                quotes,
                shapes}
\newcommand{\comment}[1]{}
\newcommand{\vsa}{\vspace*{-0.4cm}}
\newcommand{\vsb}{\vspace*{-0.2cm}}

\newcommand\todo[1]{\textcolor{red}{#1}}

\newtheorem{theorem}{Theorem}[section]

\newtheorem{lemma}[theorem]{Lemma}

\long\def\cut#1{{}}

\newif\ifIncludeEpsilon
\IncludeEpsilontrue

\def\top2{{\mbox{max2}\,}}
\title{Modeling Human-AI Team Decision Making}
\author {
	Wei Ye,\textsuperscript{\rm 1}
	Francesco Bullo, \textsuperscript{\rm 2}
	Noah Friedkin, \textsuperscript{\rm 2}
	Ambuj K Singh \textsuperscript{\rm 2}\thanks{To whom correspondence should be addressed.}
}
\affiliations {
	\textsuperscript{\rm 1} Tongji University, Shanghai 201804, China\\
	\textsuperscript{\rm 2} University of California, Santa Barbara, CA 93106, USA\\
	ambuj@cs.ucsb.edu
}

\begin{document}

\maketitle 

\begin{abstract}
AI and humans bring complementary skills to group deliberations. Modeling this group decision making is especially challenging when the deliberations include an element of risk and an exploration-exploitation process of appraising the capabilities of the human and AI agents. To investigate this question, we presented a sequence of intellective issues to a set of human groups aided by imperfect AI agents. 
%where the group had an option to reach a decision on each issue with or without AI input. %If an AI agent is invoked, then the group must decide on whether or not to accept AI agent's position on the issue, or seek a consensus based on its members' positions. Immediate feedback is given on whether the group's response is correct or not. 
A group's goal was to appraise the relative expertise of the group's members and its available AI agents, evaluate the risks associated with different actions, and maximize the overall reward by reaching consensus. We propose and empirically validate models of human-AI team decision making under such uncertain circumstances, and show the value of socio-cognitive constructs of prospect theory, influence dynamics, and Bayesian learning in predicting the behavior of human-AI groups. %We also find that groups reach consensus more easily when interacting solely with other humans and, in fact, they usually achieve better performance on average than the best member on the group. %We find that groups have difficulties interacting with AI agents, both in deciding when to seek their help and whether the AI input should be trusted. 
\end{abstract}

\section{Introduction}
%teams
%History tells us that we have always accomplished the most while working together. 
Group decision making has been ever important in organizations and missions. And now, intelligent agents have become fundamental to everyday life: assistants on mobile devices, pedagogical agents in tutoring systems, and robots collaborating with humans. As such, integrating AI agents into human groups while being aware of their complementary strengths and abilities has become vital~\cite{crandall2018cooperating,gaur2016effects,kamar2012combining,shirado2017locally,wang2016deep}. We explore models for decision making in mixed teams while building on the constructs of prospect theory and appraisal systems.
Groups/teams frequently encounter decisions involving varying amounts of risk and reward. Cumulative Prospect Theory~\cite{tversky1992advances} (formerly Prospect Theory~\cite{kai1979prospect}) provides a basis for such decision-making by establishing that individuals make decisions based on the potential value of losses and gains among the set of available options. Cumulative Prospect Theory proposes that individuals compute an internal evaluation for each prospect that is determined by a value function and a probability weighting function. The value function is S-shaped and asymmetrical, capturing loss aversion. The probability weighting function encodes the hypothesis that individuals overreact to small probability events, but under-react to large probability events.

%appraisal dynamics
Group decision making is based on how team members appraise others. Team members' opinions of others depend on the issue and evolve over time, based on their performance. Such opinions can be represented by an appraisal network~\cite{mei2017dynamic}, or equivalently by its corresponding row-stochastic appraisal matrix. The edges of this matrix are weighted, depict trust/distrust, friendship/animosity, or more generally to what degree a team member is influenced by particular others. The eigenvector centrality of each person in an irreducible aperiodic appraisal matrix is a summary measure of each team member's relative importance in determining the team's decision making.

% description of experimental setup
In order to understand how a human-AI team reaches a decision in uncertain environments, we carried out cyber-human experiments in which a team is asked to answer a sequence of questions. The task is to answer intellective questions from different categories such as history, science and technology, etc. Every team consists of four members each with access to its own AI agent. The AI agents give the correct answer at a fixed probability (unknown to the team members); however, the baseline probabilities vary across the agents. Each question is answered in four timed phases. In the first phase, every team member records his/her individual response for the question. In the second phase, the response of every team member is displayed on the screen and a chat plugin (the only communication channel) is enabled for communication. Team members then record their choices and decide whether or not to use an AI agent (and which AI agent to use) in an optional third phase. In the fourth and final phase, the team submits an answer. The correct answer to each question is displayed at the end of each round. Note that if the group has relied on the incorrect response of an AI agent, then the group's trust in that agent (and their other available AI agents) may be eroded. After every five questions an influence survey is presented. This is filled in by each team member to record the relative direct influence of each of his/her teammates on his/her opinion. The team members are also asked to rate the accuracy of all four AI agents based on their interactions with them. %A self-knowledge analysis survey is carried out at the beginning and end of the experiment. 

%Model
The first step of a team's decision making consists of deliberating and choosing one of the multiple choice options or choosing to consult an AI agent. In case a team chooses to consult an AI agent, the second decision making step consists of integrating the agent's answer with the team's choices and reporting a final answer. Each possible action in a decision making step is associated with a probability of success and a reward. From the viewpoint of Prospect Theory, each action is a gamble and the team needs to choose between multiple gambles. We observe the actions by the teams but not the perceived probability of success, probability weighting function, or the valuation function. 

We investigate and compare four models that predict the decisions of the teams. The first two models capture the appraisal process in a team, while the last two models capture the appraisal process as well as decision making under risks. We outline these four models below.

The first model, {\bf NB} (Naive Bayes), captures the accuracy of a human/AI-agent using a beta distribution that is updated at each round (after observing whether it was correct or incorrect) using Bayes rule. A Naive Bayes assumption is used to integrate the responses of the human/AI-agents and obtain the probability of using an option or an agent. %\textcolor{blue}{NEF Does the term agent always refer to an AI?}
The second model, {\bf CENT} (Centrality), integrates individual responses through an interpersonal influence system. The probability of the team choosing an option is computed as the sum of the eigenvector centrality values of each individual choosing that option. The assumption underlying this computation is always satisfied in our data. A similar weighting process is used to integrate the team's evaluation of the AI-agents. The third model, {\bf PT-NB} (Prospect Theory coupled with Naive Bayes), uses Prospect Theory to analyze the actions of the team as a set of gambles. The probabilities of success and reward of each gamble are computed as in the model {\bf NB}. The team chooses among these gambles based on Prospect Theory parameters of the team (learned through an initial training sequence). The final model, {\bf PT-CENT} (Prospect Theory coupled with Centrality), again uses Prospect Theory to analyze the actions of the team as a set of gambles. The probabilities of success and reward of each gamble are computed as in the model {\bf CENT}. A team again chooses among these gambles based on Prospect Theory. 

We measure the accuracy of the above four models using our experimental data. This measurement is non-trivial since only the team's chosen action is observed and not the team's detailed preferences for each of the actions. We resolve this difficulty by considering the set of probability distributions (on the actions) over which the chosen action is dominant and computing the minimum distance (based on cross-entropy loss) over this set to the model's predicted probability distribution.

\section{Related Work}

%Research on group decision-making indicates that it cannot be understood by studying the components (individuals and networks) in isolation. 
Group performance is not simply a sum of individual performance, but ruled by patterns of interactions, influence, and other relationships among group members. We know that transactive memory systems (TMS)~\cite{DMW:87}) are activated in group members' levels of expertise and potential contributions to tasks are appraised. We know that interpersonal influence systems
%, with various weighted digraph structures of $i \rightarrow j$ arcs of accorded influence, including $i \rightarrow i$ self-weight loops, 
are automatically generated in groups~\cite{NEF-ECJ:11}.  And we know that groups' opinions on multidimensional issues are generally constrained to a decision space of feasible positions that is the group's convex hull of initial displayed positions~\cite{NEF-WM-AVP-FB:17r}. %And we know that groups' influence centralities evolve along their sequence of considered issues \cite{NEF-PJ-FB:14n}.
%\textcolor{red}{fix citation} %Scientists have made significant headway in understanding collective team behavior of humans. We have developed an understanding of natural social processes of teams on single issues and sequences of issues; we  understand the implications of these social processes for team performance on objective measures of performance, and their emergent effects given the characteristics of teams and the actors in them; and are able to offer a set of empirically vetted social devices for improving team performance.
%We have some understanding of how truth wins in groups dealing with issues that have secure foundations for the determining the truth, but in many important issues such foundations are not available. In such cases, group decisions are subject to regrets based on cognitive biases and limited exploration  of alternatives~\cite{ILJ:72} that lead to faulty conclusions.  

An excellent survey of human-AI teaming has been put forth in a recent paper~\cite{pmid33092417}. 
%A related review~\cite{pmid32863679} discusses the implications of leadership in human-AI teams. 
\cite{DMBPMM18} discusses
the mechanisms for enhancing teamwork in human-AI teams and outlines the critical scientific questions that must be addressed to enable this vision. %\cite{SEEBER2020103174} outlines a research agenda for exploring the potential risks and benefits of human-AI teams.

% subject to retrosctive ]faulty conclusions  The findings do not secure The Specifically, we have understood conditions leading a team decision process to faulty conclusions as documented in the literature on groupthink~) and prevalence of individuals. 

%Various strategies been proposed to mitigate the probability of decision regrets. These include the Delphi system method~\cite{CCH-BAS:07} and its many variations, superforecasting methods~\cite{PET-DG:16}, the surprising popularity (SP) method~\cite{DP-HSS-JMC:17}). These methods invoke group procedural rules on the iterative interaction of experts and the form in which their opinions can be displayed.  %We will review these methods in Section~\ref{sec:decision-estimation-forecasting} below. \textcolor{red}{the section label is missing}
%\subsection{Teams of agents:}

Group constructs have also been proposed in machine learning---experts, weak learners, crowd-sourced workers---to achieve goals that no single individual can accomplish on its own. In the case of boosting~\cite{RES:90}, one can obtain a ``strong learner'' that is able to predict arbitrarily accurately based on an ensemble of ``weak learners'' whose predictions are slightly better than random guessing. In the case of ``learning from expert advice''~\cite{NCB-YF-DH-DPH-RES-MKW:97}, an algorithm works with a group of $K$ arbitrary ``experts'' who give daily ``stock predictions'' and who perform nearly as well as the ``expert'' that has the best ``track record'' at any given time. It is an iterative game in which in each iteration the ``player'' must make a decision and the experts with the best track record may change over time. The ``Multi-armed Bandits'' (MAB) problems~\cite{PA-NCB-PF:02} can be thought of as a variant of the problem of ``learning from expert advice'' in which a ``player'' can only observe the payoff of the ``expert'' at each iteration. %An algorithm makes online decisions that balance ``exploration'' and ``exploitation'', so as to maximize the cumulative payoffs. %One can conceptualize each team as an arm that is chosen in response to an arriving object.
In the case of ``crowd sourcing''~\cite{PW-SB-PP-SJB:10}, an algorithm  aggregates the inputs of a large group of unreliable ``participants'', evaluates each ``participant'',  and then infers the ground truth. %Here the assumption is an unbiased distribution of inputs.

Humans and AI are clearly different in
their cognitive and processing capabilities~\cite{MC:14}. Groups with AI involvement should be designed so that the  raw computational and search power of computers for state-space reduction can be combined with group  inductive reasoning, especially in uncertain environments. What is the optimal group-AI design for a given decision? This is a question pervading all kinds of groups that oriented to specific types of issues. % but becomes especially acute for human-agent teams. 
Taxonomies and ontologies for characterizing group decision-making %autonomy in tasks 
have been defined~\cite{TBS-WLV:78,MRE-DBK:99,JR:87} in order to investigate the optimal composition of groups. The behavior of groups with AI involvement must be observable and predictable.
%for effective human-agent teams. 
This is challenging in complex uncertain environments. %because humans and agents reach non-optimal decisions in different manners: 
While groups often adopt satisficing strategies~\cite{HAS:57,NEF-WM-AVP-FB:17r}, AI utilizes search space reduction strategies such as limited look-ahead, constraint relaxation, and heuristics. Both groups and AI are subject to bias and faulty information: groups by their members' beliefs and AI by the available data and training protocols. Since observability and predictability have ramifications on the level of trust~\cite{KS-PJH-DW:07}, groups with AI involvement must have confidence that the behavior of their AI %~\cite{stubbs2007autonomy} there is a need for humans and agents to explain their actions, or else performance can suffer~\cite{stubbs2007autonomy}. Establishing 
is consistent with an acceptable common ground whatever the displayed initial beliefs of the group's members might be %(or a shared mental model) between the team members while working on tasks is essential
~\cite{ARM-KES-AWE-JSM:18,SL-WS-TM:16,JC-MB:14,
  QKT-MEL-JV-MBM-STG-DSB:16:16, JOG:17,
  ARM-JSM-BJL-JRL-DJ-KTL-JT-ES-BMS-WN-KMD:18}. 
  %The importance of engineering high performance group analysis and decision making cannot be understated. %Famous high-impact mistakes in human deliberative teams include the classic case-studies by Janis~\cite{ILJ:72}: the Bay of Pigs disaster in 1961 and the Japanese attack on Pearl Harbor in 1941. A recent notable example was the failure of US intelligence analysts to correctly determine the absence of WMDs in the lead up to the Iraq war. 
%This project will feature a comprehensive approach to the modeling, design and validation of human-AI teams. 

Theory of mind (ToM) \cite{PW1978, BS2011, CMCS2020, OHS2021, rabinowitz2018machine} broadly refers to humans’ ability to represent the mental states of others, including their desires, beliefs, and intentions. This ability to attribute mental states to others is a key component of cognition that needs to be incorporated into AI in order to model and interact with humans. 

Human teams can display magnified cognitive capacity and unique cognitive abilities that emerge from the interaction between the team members. Collective intelligence refers to a team’s ability to produce intelligence and behaviors beyond the individual~\cite{AWW-CFC-AP-NH-TWM:10}. How to integrate AI into human teams in order to produce cognitive abilities that go beyond the individual or the group of humans is an important question~\cite{bansal2019beyond}.

Research from psychology suggests that people process uncertainty and information in general using dual processes: an implicit (automatic), unconscious process and an explicit (controlled), conscious process~\cite{SC-YT:99}. The second process is encoded by analytic algorithms, rules, and reasoning systems, and can be modeled equally well for humans and AI agents. It is the first implicit automatic system and its interaction with the explicit system that is harder to model in humans, and poses challenges for a theoretical understanding of mixed human-AI teams.

Uncertainty itself can be separated into two kinds: aleatoric and epistemic~\cite{ADK-OD:09,CRF-GU:11}. Aleatoric uncertainty refers to the notion of randomness (as in coin flipping): the variability in the outcome of an experiment that is due to inherently random effects. %There is no way to reduce this kind of uncertainty and any performance measure has to consider multiple outcomes. As opposed to aleatoric uncertainty, 
Epistemic uncertainty refers to uncertainty caused by a lack of knowledge of decision makers. This uncertainty can in principle be reduced by a proper recognition of expertise on teams and protocols that reveal explanations on why a fact may be true.

Both aleatoric and epistemic uncertainty require teams to deal with decisions involving varying amounts of risk and reward under conditions that are not completely rational. The most successful behavioral model of risky decision making is prospect theory~\cite{DK-AT:79, AT-DK:92}. According to it, individuals make decisions based on the potential value of losses and gains among the set of available options. It proposes that individuals compute an internal evaluation for each prospect that is determined by a value function and a probability weighting function. The value function is S-shaped and asymmetrical, capturing loss aversion. The probability weighting function encodes the hypothesis that individuals overreact to small probability events, but under-react to large probability events. The theory deviates from its rational competitor, expected utility theory~\cite{PCF-CSF:70}, which assumes that people evaluate the outcome of a decision in terms of the expected reward, independent of any cognitive biases (such as risk aversion). Other recent theories explaining individual choices under risk/uncertainty include dynamic decision models~\cite{PDW2015} such as dynamic field theory~\cite{pmid8356185}.

\section{Proposed Models of Decision Making}

\subsection{Decision Tasks}\label{sec:dt}
%\textcolor{blue}{NEF: The rule is that the experimental design must be described in sufficient detail to allow a replication of the experiment. Please carefully review the following text. }

A team in our experiments consists of four humans and four AI agents. The accuracy of each AI agent is fixed during an experiment and ranges between 0.6 and 0.9. Humans are informed that the accuracy of each AI agent is at least 0.5. The reward for a correct answer $c_1$ is 4, the penalty for an incorrect answer $c_2$ is 1, and the penalty for consulting an AI agent $c_3$ is 1. %\textcolor{blue}{This incentive design encourages an effort by the team to do what? To get an agreed upon answer without or with an AI agent? (Response: to design gambles with differing rewards in Decision Task 1)} \textcolor{red}{This is not really an answer to my question. The incentive design is the same for all questions. It encourages the group to obtain a correct answer, and discourages a group from using an AI that has provided an incorrect answer. If the group is failing to get the correct answer, then a search for the most competent AI is encouraged.}  
Multiple choice questions are posed sequentially to a team. After answering every five questions, each team member reports the relative influence of each member, which collectively generates a row stochastic weight matrix $W$ (and the influence network) for the team. Each team member also rates the accuracy of the AI-agents after every five questions.

There are two sequential decision tasks in our experiment:
% \begin{description}
%     \item[Decision Task 1. Integration of human responses and decision on the use of AI agents] A team first needs to integrate the decisions of the team members into a group response, and decide whether it needs to utilize an AI agent to help answer the posed question correctly. And if an AI agent is to be used, the team needs to decide which of the four AI agents to use.
%     \item[Decision Task 2. Integration of AI agent and human responses] If an AI agent is used, then the team needs to integrate its response with the prior human responses into a group response.
% \end{description}

\begin{itemize}
\item \textbf{Decision Task 1. Integration of human responses and decision on the use of AI agents.}
A team first needs to integrate the decisions of the team members into a group response, and decide whether it needs to utilize an AI agent to help answer the posed question correctly. And if an AI agent is to be invoked, the team needs to decide which of the four AI agents to use.
\item \textbf{Decision Task 2. Integration of AI agent and human responses.}
If an AI agent is consulted, then the team needs to integrate its response with the prior human responses into a group response.

\end{itemize}

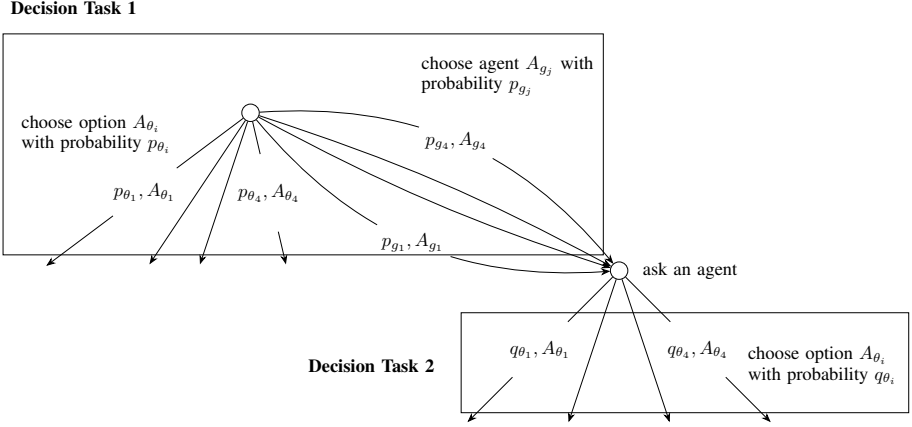
\begin{figure*}[!htb]
	\centering
	\scalebox{0.7}{
		\begin{tikzpicture}
		\begin{scope}
		\node (A) at (3,3) [circle,draw] {};
		\node (C) at (-1,0) [] {};
		\node (D) at (1,0) [] {};
		\node (E) at (2,0) [] {};
		\node (F) at (3.7,0) [] {};
		
		\node (B) at (10,0) [circle,draw] {};
		\node (G) at (7,-3) [] {};
		\node (H) at (9,-3) [] {};
		\node (I) at (11,-3) [] {};
		\node (J) at (13,-3) [] {};
		\end{scope}
		
		\begin{scope}[>={Stealth[black]},
		every node/.style={fill=white,opacity=1,circle},
		every edge/.style={draw,bend right=0}]
		\node[text width=3.5cm] at (8,3.7) {choose agent $A_{g_j}$ with probability $p_{g_j}$};
		\draw[->] (A) edge [bend left=5] (B);
		\draw[->] (A) edge [bend left=27] node {$p_{g_4},A_{g_4}$} (B);
		\draw[->] (A) edge [bend right=5] (B);
		\draw[->] (A) edge [bend right=27] node {$p_{g_1},A_{g_1}$} (B);
		\end{scope}
		
		\begin{scope}[>={Stealth[black]},
		every node/.style={fill=white,opacity=1,circle},
		every edge/.style={draw,bend right=0}]
		\node[text width=3.1cm] at (0.0,5) {\textbf{Decision Task 1}};
		\node[text width=3.1cm] at (.2,2.6) {choose option $A_{\theta_i}$ with probability $p_{\theta_i}$};
		\draw[->] (A) edge node {$p_{\theta_1},A_{\theta_1}$} (C);
		\draw[->] (A) edge (D);
		\draw[->] (A) edge (E);
		\draw[->] (A) edge node {$p_{\theta_4},A_{\theta_4}$} (F);
		
		\draw [draw=black,even odd rule] (-1.7,4.5) rectangle (9.7,0.3);
		\node[text width=3.5cm] at (12.2,0) {ask an agent};
		\node[text width=3.1cm] at (14,-1.8) {choose option $A_{\theta_i}$ with probability $q_{\theta_i}$};
		\draw[->] (B) edge node {$q_{\theta_1},A_{\theta_1}$}  (G);
		\draw[->] (B) edge (H);
		\draw[->] (B) edge (I);
		\draw[->] (B) edge node {$q_{\theta_4},A_{\theta_4}$} (J);
		\node[text width=3cm] at (5.6,-1.8) {\textbf{Decision Task 2}};
		\draw [draw=black,even odd rule] (7,-0.8) rectangle (15.5,-2.7);
		\end{scope}
		\end{tikzpicture}
	}
	\vsa 
	\caption{Decision making in the human-AI teams. The first decision task involves choosing one of the four options presented with each question or choosing one of the agents for help. The second decision task integrates the agent's response with the earlier individual choices.}
	\label{fig:decisionTasks}
	\vsa \vsb 
\end{figure*}

Figure~\ref{fig:decisionTasks} shows the two decision tasks. The first decision task consists of eight possible actions: the first four actions $A_{\theta_1}, \ldots, A_{\theta_4}$ correspond to the decision of using one of the four choices to the posed question without consulting an agent, and the next four actions $A_{g_1}, \ldots, A_{g_4}$, correspond to the decision of using an AI agent. Action $A_{\theta_i}$ is associated with probability $p_{\theta_i}$ while action $A_{g_i}$ is associated with probability $p_{g_i}$. For the second decision task, the team needs to integrate the answer of an agent with the answers of the human team members and decide on one of the four choices. In this case, the task consists of four actions: $A_{\theta_1}, \ldots, A_{\theta_4}$, which are associated with probabilities $q_{\theta_1}, \ldots, q_{\theta_4}$, respectively. For a team to be successful, it needs to reach consensus on which action to take in the first decision task; furthermore, if one of the agents is chosen, then the team needs to achieve consensus on one of the four actions in the second decision task.

We next present a sequence of models that explain the decision making of a team. The first model uses Bayes rules to combine the individual choices into the team's response. The second model achieves this integration through a weighting mechanism based on the eigenvector centralities of the individuals. The last two models utilize Prospect Theory to reflect how teams combine appraisal with decision making in risky environments. 

\subsection{Model NB: Integration through Bayes Rule}

Model {\bf NB} uses Bayes rule to integrate human-AI choices into a team response. We explain the model by considering first decision task depicted in Figure~\ref{fig:decisionTasks}. The modeling can be broken down into three steps:  Step 1 considers the individual responses made by individuals, Step 2 considers the choice of an agent, and Step 3 integrates the choices made in the previous two steps into a model for the eight actions in Figure 1.

\noindent {\bf Step 1:}
Let $\Theta$ be the random variable corresponding to a team's collective response. It ranges over the four choices to the posed question represented by $\theta_i, i = 1 \ldots 4$. Let $\theta^*$ be the correct choice. Let random variable $R_i$ be the response of individual $i$.  Model {\bf NB} predicts the probability of the team reaching the decision on choice $\theta_k$ as follows: 
 \begin{multline}
     P(\Theta = \theta_k | R_i, i = 1\ldots 4)\\
 = \frac{P(R_i, i = 1\ldots 4 | \Theta = \theta_k) \times P(\Theta = \theta_k)}{P(R_i, i = 1\ldots 4)}
 \end{multline}

The above equation can be reduced by assuming that the individual choices are independent given the team's choice:
% \begin{equation}
% \begin{aligned}
% p'(\theta_k) &= P(\Theta = \theta_k | R_i, i = 1\ldots 4)\\
% &= \frac{
% (\Pi_{i=1}^4 P(R_i| \Theta = \theta_k)) \times P(\Theta = \theta_k)
% }{P(R_i, i = 1\ldots 4)}
% \end{aligned}
% \label{eq:bayes}
% \end{equation}
\vsb 
 \begin{multline}
p'_{\theta_k} = P(\Theta = \theta_k | R_i, i = 1\ldots 4)\\
= \frac{
(\Pi_{i=1}^4 P(R_i| \Theta = \theta_k)) \times P(\Theta = \theta_k)
}{P(R_i, i = 1\ldots 4)}
\label{eq:bayes}
 \end{multline}
 \vsb 

When $R_i = \theta_k$, the term $P(R_i | \Theta = \theta_k)$ is assumed to equal $E[\mathbf{1}\lbrace R_i= \theta^*\rbrace]$, the expected performance of individual $i$. Otherwise, $P(R_i | \Theta = \theta_k) = \frac{1-E[\mathbf{1}\lbrace R_i= \theta^*\rbrace]}{3}$.
In other words, a team deciding the same as an individual's choice assumes that the individual is correct and a team deciding differently from an individual assumes that the individual is incorrect. These assumptions imply that the group discovers the truth only if some individual suggests it. The value $E[\mathbf{1}\lbrace R_i= \theta^*\rbrace]$ is defined by a beta distribution with two parameters that are initialized to one (leading to a uniform prior over the range [0,1]), and updated after each response by an individual. The prior probabilities $P(\Theta = \theta_k)$ are assumed to be $0.25$, and the term in the denominator of Eq.~\ref{eq:bayes} is found by normalization. Note that $\mathbf{1}$ is the indicator function.

\noindent {\bf Step 2:}
The expected performance of agent $i$ is again modeled using a beta distribution. The initial values of the two parameters are chosen so that we have a uniform prior over [0.5,1] to account for the fact that the accuracy of each agent is declared at the outset to be at least 0.5. The parameters are updated following each round in which an agent is consulted and its accuracy is observed. The probability of choosing agent $g_i$ is $p'_{g_i} = \frac{E[\mathbf{1}\lbrace G_i= \theta^*\rbrace]}{\Sigma_j E[\mathbf{1}\lbrace G_i= \theta^*\rbrace]]}$.

\noindent {\bf Step 3:}
Based on the probability of each action (choosing an option or an agent), we compute the expected reward for each possible action. The reward for choosing option $\theta_i$ is $x_{\theta_i} = c_1 \times p'_{\theta_i} - c_2 \times (1-p'_{\theta_i})$. And the reward for consulting agent $i$ is $x_{g_i} = (c_1-c_3) \times p'_{g_i} - (c_2+c_3) \times (1-p'_{g_i})$. We use the softmax function to transform $x_{\theta_i}$ and $x_{g_i}$ into probability values:
\vsb 
\begin{gather}
p_{\theta_i} = \frac{e^{x_{\theta_i}}}{\Sigma_j e^{x_{\theta_j}} + \Sigma_j e^{x_{g_j}}}, \\
p_{g_i} = \frac{e^{x_{g_i}}}{\Sigma_j e^{x_{\theta_j}} + \Sigma_j e^{x_{g_j}}}.
\end{gather}
This concludes the modeling of the first decision task. Next, we consider how model {\bf NB} explains the second decision task. Let $g_j$ be the agent that the team consulted. We integrate the response $G_j$ of the agent as follows:
% \begin{equation}
% \begin{aligned}
% q'(\theta_k) &= P(\Theta = \theta_k | R_i, i = 1\ldots 4, G_j)\\
% &= \frac{
% (P(G_j | \Theta = \theta_k)\Pi_{i=1}^4 P(R_i | \Theta = \theta_k)) \times P(\Theta = \theta_k)
% }{P(R_i, i = 1\ldots 4, G_j)}
% \end{aligned}
% \label{eq:bayes2}
% \end{equation}
\vsb 
 \begin{multline}
q'_{\theta_k} = P(\Theta = \theta_k | R_i, i = 1\ldots 4, G_j)\\
= \frac{
(P(G_j | \Theta = \theta_k)\Pi_{i=1}^4 P(R_i | \Theta = \theta_k)) \times P(\Theta = \theta_k)
}{P(R_i, i = 1\ldots 4, G_j)}
\label{eq:bayes2}
\end{multline}
%\vsb 
When $G_j = \theta_k$, the term $P(G_j | \Theta = \theta_k)$ is assumed to equal $E[\mathbf{1}\lbrace G_j = \theta^* \rbrace]$, the expected performance of agent $j$. Otherwise,  $P(G_j | \Theta = \theta_k) = \frac{1-E[\mathbf{1}\lbrace G_j = \theta^* \rbrace]}{3}$. $P(R_i | \Theta = \theta_k)$ is defined similarly (and as in Eq.~\ref{eq:bayes}). Based on the above probabilities, we again compute the expected reward for each possible action. The reward for choosing option $\theta_i$ is $y_{\theta_i} = (c_1-c_3) \times q'_{\theta_i} - (c_2+c_3) \times (1-q'_{\theta_i})$. We again use the softmax function to transform $y_{\theta_i}$ into probability values:
\vsa 
\begin{equation}
q_{\theta_i} = \frac{e^{y_{\theta_i}}}{\Sigma_j e^{y_{\theta_j}}}
\end{equation}

\subsection{Model CENT: Integration by Interpersonal Influence System}
\begin{figure}[!htb]
	\centering
	\scalebox{0.9}{\begin{tikzpicture}
		\begin{scope}
		\node (A) at (0,3) [circle,draw] {$H_1$};
		\node (A1) at (0,3.7) [circle] {$\delta_1$};
		\node (B) at (2,3) [circle,draw] {$H_2$};
		\node (B1) at (2,3.7) [circle] {$\delta_2$};
		\node (C) at (4,3) [circle,draw] {$H_3$};
		\node (C1) at (4,3.7) [circle] {$\delta_3$};
		\node (D) at (6,3) [circle,draw] {$H_4$};
		\node (D1) at (6,3.7) [circle] {$\delta_4$};
		\node (E) at (0,0) [rectangle,draw] {$A_1$};
		\node (E1) at (0,-0.6) [circle] {$\pi_1$};
		\node (F) at (2,0) [rectangle,draw] {$A_2$};
		\node (F1) at (2,-0.6) [circle] {$\pi_2$};
		\node (G) at (4,0) [rectangle,draw] {$A_3$};
		\node (G1) at (4,-0.6) [circle] {$\pi_3$};
		\node (H) at (6,0) [rectangle,draw] {$A_4$};
		\node (H1) at (6,-0.6) [circle] {$\pi_4$};
		\end{scope}
		
		\begin{scope}[>={Stealth[black]},
		every node/.style={fill=white,circle},
		every edge/.style={draw,bend right=0}]
		\draw[<->] (A) edge (B);
		\draw[<->] (B) edge (C);
		\draw[<->] (C) edge (D);
		\draw[<->] (A) edge [bend left=40] (C);
		\draw[<->] (A) edge [bend left=40] (D);
		\draw[<->] (B) edge [bend left=40] (D);
		\draw [draw=black,fill=black,opacity=0.2,even odd rule] (-0.5,4.5) rectangle (6.5,2.5);
		\end{scope}
		
		\begin{scope}[>={Stealth[black]},
		every node/.style={fill=white,circle},
		every edge/.style={draw,bend right=0}]
		\draw[->] (A) edge node {$\pi_{1,1}$} (E);
		\draw[->] (A) edge (F);
		\draw[->] (A) edge (G);
		\draw[->] (A) edge (H);
		%[fill=white,opacity=0,text opacity=1]
		\draw[->] (B) edge node {$\pi_{2,1}$}  (E);
		\draw[->] (B) edge (F);
		\draw[->] (B) edge (G);
		\draw[->] (B) edge (H);
		\draw[->] (C) edge node {$\pi_{3,1}$} (E);
		\draw[->] (C) edge (F);
		\draw[->] (C) edge (G);
		\draw[->] (C) edge (H); 
		\draw[->] (D) edge node {$\pi_{4,1}$}  (E);
		\draw[->] (D) edge (F);
		\draw[->] (D) edge (G);
		\draw[->] (D) edge (H);
		\end{scope}
		
		\end{tikzpicture}
	}
	\vsa 
	\caption{The appraisal networks. The shaded box is the appraisal matrix between the individuals on the team. The eigenvalues of the matrix are denoted by $\delta$. These individuals also appraise the agents. The $\pi$ values denote these appraisals.}
	\vsa 
	\label{fig:agentRatings}
\end{figure}
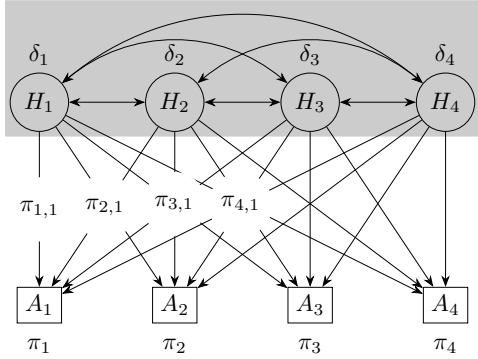

This model integrates individual responses through an interpersonal influence system. Each individual in the experiment is asked to rate the influence of other members in their team after every five questions. This leads to an influence network $N$ (as shown in Figure~\ref{fig:agentRatings}) whose centralities are used to weigh the choices of the team members. Consider Decision Task 1 (Figure~\ref{fig:decisionTasks}) first. The modeling is again done in three steps. 

\paragraph{Step 1:}
The probability of the team choosing option $\theta_k$ is modeled as the sum of the eigenvector centrality values in network $N$ of each individual who has chosen option $\theta_k$. If the centrality value of individual $i$ is $\delta_i$ then we obtain the following for the probability of choosing option $\theta_k$.
\begin{equation}
p'_{\theta_k} = P(\Theta = \theta_k | R_i, i = 1\ldots 4)
= \frac{\Sigma_i \mathbf{1}_{\theta_k}(R_i) \times \delta_i}{P(R_i, i = 1\ldots 4)}
\label{eq:cent}
\end{equation}

\paragraph{Step 2:}
This step uses a weighting process to model the team's collective appraisal of the agents. Let the $i$-th individual's appraisal of the $j$-th agent be $\pi_{ij}$. Note that the values $\pi_{ij}$ are collected during the experiment after every five questions. Let the eigenvector centrality values of individual $i$ be $\delta_i$ as before. We model the collective appraisal of agent $j$ as $\pi_j$ and the probability of choosing agent $g_j$ as $p'_{g_j}$:
\begin{gather}
\pi_j = \Sigma_i \delta_i \pi_{i,j} \\
p'_{g_j} = \frac{\pi_j}{\Sigma_j \pi_{j}}.
\end{gather}

\paragraph{Step 3:}
As in Step 3 of {\bf NB}, we first compute the reward for each possible action and then transform all the rewards to probabilities using the softmax function. 

This concludes the modeling of the first decision task. For the second decision task, we need to integrate the response $G_j$ of the consulted agent $j$ into the collective choice of the team. This integration now (as opposed to Decision Task 1) needs to consider the choice made by the agent and the choices made by the team prior to consultation with the agent. We utilize a parameter $w$ for this integration: $w$ weighs the influence placed in the agent's response as compared to the influence placed in the initial human responses. This (global) parameter is learned by a training sample of teams. We have the following resulting probability $q'_{\theta_k}$ of choosing option $\theta_k$ in Decision Task 2.
% \begin{equation}
% \begin{aligned}
% q'(\theta_k) &= P(\Theta = \theta_k | R_i, i = 1\ldots 4, G_j)\\
% &= \frac{ \mathbf{1}_{\theta_k}(G_j) \times \pi_j \times w + \Sigma_i \mathbf{1}_{\theta_k}(R_i) \times \delta_i \times (1-w)}{P(R_i, i = 1\ldots 4, G_j)}
% \end{aligned}
% \label{eq:cent2}
% \end{equation}
\vsa 
\begin{multline}
q'_{\theta_k} = P(\Theta = \theta_k | R_i, i = 1\ldots 4, G_j)\\
= \frac{ \mathbf{1}_{\theta_k}(G_j) \times \pi_j \times w + \Sigma_i \mathbf{1}_{\theta_k}(R_i) \times \delta_i \times (1-w)}{P(R_i, i = 1\ldots 4, G_j)}
\label{eq:cent2}
\end{multline}
Based on these probabilities, we compute the expected rewards and derive the probabilities of choosing the four actions as in model {\bf NB}.

\subsection{Model PT-NB: Integration through Prospect Theory and Naive Bayes}

This model uses Prospect Theory to analyze the actions of the team as a set of gambles. For the first decision task, the estimation of probabilities $p'_{\theta_i}$,  $p'_{g_i}$ (Steps 1 and 2) is exactly as in model {\bf NB}. Step 3 now utilizes Prospect Theory parameters to integrate them and obtain the probabilities for the eight actions.

We model the choices made by the team using five Prospect Theory parameters ($\alpha, \beta, \lambda, \gamma^+, \gamma^-$). $\alpha=\beta \in [0,1]$, $\gamma^{+/-} \in [0,1]$, and $\lambda \in [0,10]$. The gamble for option $\theta_i$ is $\{(c_1, p'_{\theta_i}), (c_2, 1-p'_{\theta_i})\}$ and the gamble for the use of agent $g_i$ is $\{(c_1-c_3), p'_{g_i}), (c_2+c_3, 1-p'_{g_i})\}$. The corresponding rewards $x_{\theta_i}$ and $x_{g_i}$ are computed as follows: 
 \begin{equation}
 \begin{aligned}
 x_{\theta_i}&= c_1^{\alpha} \times \exp(-(\log \frac{1}{p_{\theta_i}})^{\gamma^+})\\ 
 &+ \left(-\lambda|c_2|^{\beta} \times \exp({-(\log \frac{1}{1-p_{\theta_i}})^{\gamma^-}})\right) \\
 x_{g_i}&= (c_1-c_3)^{\alpha} \times \exp(-(\log \frac{1}{p_{g_i}})^{\gamma^+}) \\
 &+ \left(-\lambda|c_2 + c_3|^{\beta} \times \exp({-(\log \frac{1}{1-p_{g_i}})^{\gamma^-}})\right)
 \end{aligned}
 \label{eq:PTvalue}
 \end{equation}

%\begin{gather}
%x_{\theta_i}= c_1^{\alpha} \times \exp(-(\log \frac{1}{p_{\theta_i}})^{\gamma^+}) 
%+ \left(-\lambda|c_2|^{\beta} \times \exp({-(\log \frac{1}{1-p_{\theta_i}})^{\gamma^-}})\right),\\
%x_{g_i}= (c_1-c_3)^{\alpha} \times \exp(-(\log \frac{1}{p_{g_i}})^{\gamma^+}) 
%+ \left(-\lambda|c_2 + c_3|^{\beta} \times \exp({-(\log \frac{1}{1-p_{g_i}})^{\gamma^-}})\right).
%\label{eq:PTvalue}
%\end{gather}
Finally, we use the softmax function to transform these rewards into probability values $p_{\theta_i}$ and $p_{g_i}$.

This concludes the modeling of the first decision task. The computations for the second decision task are similar. We use the probabilities $q'_{\theta_k}$ from model {\bf NB} and the Prospect Theory parameters to compute the rewards $y_{\theta_i}$ and these are converted into probabilities $q_{\theta_i}$ using the softmax function. Note that the gambles for all four options involve the same rewards and this will result in a similar model as {\bf NB}.

\subsection{Model PT-CENT: Integration through Prospect Theory and Interpersonal Influence System}

The development of this model is similar to {\bf PT-NB} except that the model {\bf CENT} is now used for computing the probabilities $p'_{\theta_i}$, $p'_{g_i}$ and $q'_{\theta_i}$. Using the Prospect Theory parameters, we obtain the rewards $x_{\theta_i}$, $x_{g_i}$ and $y_{\theta_i}$. Finally, we use the softmax function to obtain the probabilities $p_{\theta_i}$ and $p_{g_i}$ for the eight actions in Decision Task 1 shown in Figure~\ref{fig:decisionTasks}, and the probabilities $q_{\theta_i}$ for the four actions in Decision Task 2 shown in Figure~\ref{fig:decisionTasks}. Again, since the gambles for the four options involve the same rewards, this will result in a similar model as {\bf CENT}.

% Our main experimental results are about how well the proposed models predict the dynamics of decision making in human-AI teams. In order to measure this correspondence, we first need to define a measurement criterion, or a loss function, that measures the correspondence between the model and the experimental data. 

\section{Evaluation of Models}

\subsection{Experimental Details}\label{sec:exp_design}
%\textcolor{blue}{NEF: The rule is that the experimental design must be described in sufficient detail to allow a replication of the experiment. Please carefully review the following text. }

%A team in our experiments consists of four humans and four AI agents. The accuracy of each AI agent is fixed during an experiment and ranges between 0.6 and 0.9. Humans are informed that the accuracy of each AI agent is at least 0.5. The reward for a correct answer $c_1$ is 4, the penalty for an incorrect answer $c_2$ is 1, and the penalty for consulting an AI agent $c_3$ is 1. %\textcolor{blue}{This incentive design encourages an effort by the team to do what? To get an agreed upon answer without or with an AI agent? (Response: to design gambles with differing rewards in Decision Task 1)} \textcolor{red}{This is not really an answer to my question. The incentive design is the same for all questions. It encourages the group to obtain a correct answer, and discourages a group from using an AI that has provided an incorrect answer. If the group is failing to get the correct answer, then a search for the most competent AI is encouraged.}  
We worked with 30 teams, and 45 questions were sequentially posed to each team through the POGS~\cite{kim2017makes} interface, a software system designed for conducting team experiments. Each intellective question is coupled with four alternative answers. Each question is answered in four phases, of duration 30 seconds, 30 seconds, 15 seconds and 45 seconds, respectively. In the first phase, every team member records his/her initial individual response for the question. In the second phase, the responses of every team member are displayed on the screen and the chat plugin is enabled for communication. In the next 15 seconds, the team decides whether or not to invoke the help of an AI agent (and/or which AI agent to invoke); only one AI agent can be invoked per question. In the fourth phase, the team submits an answer. The correct answer to each question is displayed after submitting the answer. Over the sequence of 45 questions, this process generates an associated sequence of nine influence networks (denoted by a row-stochastic weight matrix $W$). Consensus is reached by a team if each member chooses the same answer or chooses to invoke the same agent for help. If an agent is consulted then the team again needs to reach consensus on which answer to report. If consensus is not reached, then the task has failed and the team's response is deemed incorrect. %Usually, a consensus was not achieved during 30 sec chat period, and the team entered into the 15 sec period of deciding on whether or not use an AI without reaching a prior agreement. During the final 45 sec period an answer to the question was submitted.
%Each team member also rates the accuracy of the AI-agents after every five questions.

Our analysis is based on the $9 \times 30 =270$ occasions of team answers to the nine questions that involved a report of weight matrices. All the reported $W$ are irreducible and aperiodic structures with a unique normalized left eigenvector, which we take as the measure of each team member's relative influence centrality. (These $W$ matrices as well as the agent appraisal matrices $\pi$ do converge in our experiments.) In turn, the measure of the influence centrality of a particular initial answer is the sum of the eigenvalues of those members who chose it as their initial (pre-discussion) answer. Thus, the influence's predicted answer for the team is the multiple choice alternative that has a largest sum of eigenvalue influence centralities. %\textcolor{blue}{In section on findings, we should report findings on how many of these 270 occasions was a consensus reached prior to the 15 sec period of deciding on whether to use an AI, how many of the 270 occasions was consensus reached during the 45 sec period?}

\subsection{Results}
Our main experimental results are about how well the proposed models predict decision making. In order to measure this correspondence, we first need to define a measurement criterion, or a loss function, that measures the correspondence between the model and the experimental data.

\subsubsection{Metrics for Measuring Model Accuracy}\label{lf}
%For the first decision task, at every posed question, a team formulates a group response using the available knowledge of the team members or decides to consult one of the four AI agents. For the second decision task, at the posed question where a team is not certain about the answer and consult an agent, a team need to integrate the available knowledge of the team members and the consulted agent. In order to measure the accuracies of the models, we define the loss functions for their predictions. We consider \emph{variational loss}, which measures the cross-entropy (or the the Kullback-Leibler divergence plus an additive constant) of the model's prediction with respect to the team's decision.

In both decision tasks, each of our models generates a probability distribution over the actions (eight actions in the first decision task and four in the second). In the experimental data, we do not have the probability values the team computed, only the action chosen. As a result, we need to consider the distance between the model's predicted distribution and all possible data distributions in which the observed action is the dominant choice. We focus on measuring the smallest distance from a model to the family of possible distributions that fit the team's action. The other possibility of using an expected distance is considerably more complex since we need to assign probabilities to each feasible distribution. Let $q$ be the specific distribution produced by a model with a maxima for action $j$, and let action $i$ be the action chosen by the team. We consider all possible distributions $p$ in which the value of $p_i$ is as high as all $p_k$, and then compute the distance between the distributions to measure the correspondence. We adopt the measure of cross-entropy $H$ for measuring the correspondence. Since cross-entropy is not symmetric, this gives us two possibilities. First, the loss function $L^{(1)}$ minimizes the entropy $H(q,p)$: 
%\paragraph{Case 1: Minimize $H(q,p)$}
\begin{equation}
L^{(1)} = \min_{p\ s.t.\ \forall k:\ p_k \leq p_i} -\sum_{i=1}^n q_i\log(p_i)
\label{eq:case1_human_entropy_loss}
\end{equation}
(We have also considered the loss function $L^{(2)}$ that minimizes $H(p,q)$ with similar results.)
Instead of generating all such distributions and finding the optimal $p$, we obtain an analytical solution. The distribution $p$ that minimizes the loss function can be computed as follows:
\vsb  
\begin{align*}
i=j & \implies  && p=q \hspace*{3.5in}\\
i\neq j &\implies && p_k = 
\begin{cases} 
(1/n_H) \sum_{k \in q_{H}}q_k, \quad &\text{if }k \in q_{H} \\
q_k, \quad &\text{otherwise},
\end{cases}
\end{align*}
where $q_{H}$ consists of those indices $k$ for which $q_k \geq q_i$ and $n_H$ is the size of $q_{H}$.

Random Baseline: In order to assess the statistical significance of the models, we also compare with a baseline that makes a random choice uniformly among the possible actions at each step. For Decision Task 1, the probability of each action is 0.125 and for Decision Task 2, the probability of each action is 0.25.

\subsubsection{Learning Parameters}

For models {\bf PT-NB} and {\bf PT-CENT} in the first decision task, we use the loss functions to learn the five parameters for Prospect Theory. We use a training set of 30 questions to learn these using grid search for each team. We vary the values of $\alpha$, $\beta$ and $\gamma^{+/-}$ from 0 to 1 in steps of 0.1 and the value of $\lambda$ from 0 to 10 in steps of 1. Using these parameters, we validate the models on the remaining 15 questions in the dataset for each team. For the second decision task, since the gambles for all four options involve the same rewards, models {\bf PT-NB} and {\bf PT-CENT} are similar to models {\bf NB} and {\bf CENT}, respectively. We only report the results of the latter two models. Parameter $w$ in Equation~\ref{eq:cent2} is learned by considering a random set of 20 teams and the results are validated on the remaining 10 teams.

\subsubsection{Validation of Models}

For the first decision task in Figure~\ref{fig:decisionTasks}, the loss values of each model are given in Table~\ref{tab:loss_task1}(a). As mentioned earlier, the values are averaged over 30 teams over the last 15 questions. The Wilcoxon signed-rank test~\cite{wilcoxon1992individual} shows the results of the four models are significantly superior to that of the random model (significance level $<$ 0.01). Table~\ref{tab:loss_task1}(b) shows that the appraisal-based models {\bf NB} and {\bf CENT} perform similarly in explaining a human-AI team's decision making. The models {\bf PT-NB} and {\bf PT-CENT} also perform similarly and are superior to the previous two models (Wilcoxon signed-rank test significance level $<$ 0.01). This implies that modeling the inherent risk in decision making leads to a superior model. 

\begin{table}[!htb]
	\hspace*{\fill}
	\begin{subtable}[b]{0.45\textwidth}
		\centering
		\begin{tabular}{|l|l|}
			\hline
			Model &Loss $L^{(1)}$ \\ \hline
			{\bf NB} &$1.08 \pm 0.16$ \\
			{\bf CENT} &$1.14 \pm 0.14$ \\
			{\bf PT-NB} &$0.55 \pm 0.19$ \\
			{\bf PT-CENT} &$0.57 \pm 0.17$ \\ \hline
			{\bf RANDOM}  &$2.08 \pm 0$ \\ \hline
		\end{tabular}
		\caption{}
	\end{subtable}%
	\hfill
	\begin{subtable}[b]{0.45\textwidth}
		\centering
		\begin{tabular}{|l|l|}
			\hline
			Wilcoxon signed-rank test&$p$-value for $L^{(1)}$\\ \hline
			W({\bf NB},{\bf CENT}) &0.02\\
			W({\bf NB},{\bf PT-NB}) &$<$0.01\\
			W({\bf NB},{\bf PT-CENT}) &$<$0.01\\
			W({\bf CENT},{\bf PT-NB}) &$<$0.01\\
			W({\bf CENT},{\bf PT-CENT}) &$<$0.01\\
			W({\bf PT-NB}, {\bf PT-CENT}) &0.24\\\hline
		\end{tabular}
		\caption{}
	\end{subtable}%
	\hspace*{\fill}
	\vsa 
	\caption{(a) The loss values of ({\bf NB}, {\bf CENT}) are similar to each other and so are the loss values of ({\bf PT-NB}, {\bf PT-CENT}). The latter two models are superior suggesting that teams consider both risk/reward and appraisal in their decision making. (b) The models {\bf PT-NB} and {\bf PT-CENT} are significantly better than {\bf NB} and {\bf CENT}. (Wilcoxon signed-rank test. The null hypothesis is that the two related paired samples come from the same distribution.)}
	\vsa 
	\label{tab:loss_task1}
\end{table}

Next, we consider the second decision task in Figure~\ref{fig:decisionTasks}. Since the risk/reward of each action is the same, models {\bf PT-NB} and {\bf NB} become similar, and so do models {\bf PT-CENT} and {\bf CENT}. Thus, we only report the values for {\bf NB} and {\bf CENT}. Since the AI agents are consulted about 25.7\% of the time, there are fewer samples for this decision task. We learn the parameter $w$ by considering a random set of 20 teams and present the validation results for the remaining 10 teams in Table~\ref{tab:loss_task2}(a). The parameter $w$ is learned to be 0.9, suggesting that the teams consult the agents only when they are unsure of the choices and, as a result, and place a high degree of trust in the agent's answer. Both models are superior to the random model (Wilcoxon signed-rank test significance level $<$ 0.01).

For the second decision task (when an AI agent is consulted), we also investigate how teams integrate the responses from the AI agents in their deliberations. For this, we designed two other models for this decision task: one in which the best human response is returned (neglecting the agent response) and another in which the agent's response is returned (neglecting the human responses). Models {\bf NB-H} and {\bf CENT-H} are models of the former kind while {\bf NB-A} and {\bf CENT-A} are models of the latter kind. As can be seen from Table~\ref{tab:loss_task2}(a), models that rely only on human responses ({\bf NB-H} and {\bf CENT-H}) do not explain the team behavior, suggesting that teams do integrate the agent responses. The good behavior of the agent-only models ({\bf NB-A} and {\bf CENT-A}) as well as the high value of parameter $w$ (learned to be 0.9) suggests that teams rely much more on the agent response in this decision task. This can be explained by the fact that the teams proceed to the second decision task involving an AI agent only if they are unsure of the correct choice.

\begin{table}[!htb]
	\hspace*{\fill}
	\begin{subtable}[b]{0.45\textwidth}
		\centering
		\begin{tabular}{|l|l|}
			\hline
			Model &Loss $L^{(1)}$ \\ \hline
			{\bf NB} &$0.47 \pm 0.11$ \\
			{\bf NB-H} &$0.91 \pm 0.20$ \\
			{\bf NB-A} &$0.49 \pm 0.14$ \\ \hline 
			{\bf CENT} &$0.33 \pm 0.06$ \\
			{\bf CENT-H} &$0.95 \pm 0.17$ \\
			{\bf CENT-A} &$0.74 \pm 0.20$ \\ \hline
			{\bf RANDOM} &$1.39 \pm 0$ \\\hline
		\end{tabular}
		\caption{}
	\end{subtable}%
	\hfill
	\begin{subtable}[b]{0.45\textwidth}
		\centering
		\begin{tabular}{|l|l|}
			\hline
			Wilcoxon signed-rank test&$p$-value for $L^{(1)}$ \\ \hline
			W({\bf NB},{\bf NB-A}) &0.02\\
			W({\bf NB},{\bf NB-H}) &$<$0.01\\
			W({\bf NB-A},{\bf NB-H}) &$<$0.01\\\hline
			W({\bf CENT},{\bf CENT-A}) &$<$0.01\\
			W({\bf CENT},{\bf CENT-H}) &$<$0.01\\
			W({\bf CENT-A},{\bf CENT-H}) &$<$0.01\\\hline
		\end{tabular}
		\caption{}
	\end{subtable}%
	\hspace*{\fill}
	\vsa 
	\caption{(a) The models {\bf NB} and {\bf CENT} are the best at predicting the actions of a team. Models that rely on the human response only ({\bf NB-H} and {\bf CENT-H}) do not do as well, suggesting that the agent's response is integrated into the final decision. Models that rely on the agent only ({\bf NB-A} and {\bf CENT-A}) also perform well, suggesting that the agent's response is weighed highly in this decision task. (b) The models {\bf NB} and {\bf CENT} are significantly better than {\bf NB-H} and {\bf CENT-H}, respectively. Models {\bf NB-A} and {\bf CENT-A} are also significantly better than {\bf NB-H} and {\bf CENT-H}, respectively.}
	\label{tab:loss_task2}
	\vsa 
\end{table}

\section{Conclusions}

Our experimental setting provides a unique platform for understanding how human-AI team members learn each other's expertise and how they coordinate to solve intellective tasks in uncertain and risky environments. We proposed four models to explain the dynamics of human-AI team decision making. We find that,
although appraisal-based models {\bf NB} and {\bf CENT} perform adequately in explaining
human-AI team's decision making,
%\textcolor{blue}{There is no specification of the influence process and no basis for asserting that the reported influence weights are determined by perceived expertise. The eigenvector centrality measure is not derived from any defined mechanism of interpersonal influence. It is merely computed from the weight matrix. If it is the case that all the observed appraisal matrices are irreducible and aperiodic (or if there is only one globally reachable individual in the network) and if all groups reached a consensus during their discussions (prior to deciding on whether to use an AI, then the DeGroot influence process would be justified. But at the moment this not reported to be the case. Further note that if W is irreducible and aperiodic, then a consensus will not be reached on any one of the questions, since we are dealing with a choice between multiple discrete options.} 
the prospect theory based models {\bf PT-NB} and {\bf PT-CENT} exhibit more accurate explanations.
This finding establishes the importance of modeling the inherent risk in uncertain decision making. Our results raise a number of possibilities for future studies including the use of active AI participating agents and the use of reinforcement learning for modeling the decision making.

\bibliography{alias,references,FB,Main}

% \bibliographystyle{aaai22}
%\baselineskip

%Here you should list the contents of your Supplementary Materials -- below is an example. 
%You should include a list of Supplementary figures, Tables, and any references that appear only in the SM. 
%Note that the reference numbering continues from the main text to the SM.
% In the example below, Refs. 4-10 were cited only in the SM.    

% \clearpage

% Loss Functions\\
% Lemmas and Theorems\\
% Validation of Models using Other Loss Functions\\
% Convergence of Human-AI Appraisal Matrix\\
%Fig. 8\\
%Tables 5 to 8\\
%References \textit{(4-10)}

% For your review copy (i.e., the file you initially send in for
% evaluation), you can use the {figure} environment and the
% \includegraphics command to stream your figures into the text, placing
% all figures at the end.  For the final, revised manuscript for
% acceptance and production, however, PostScript or other graphics
% should not be streamed into your compliled file.  Instead, set
% captions as simple paragraphs (with a \noindent tag), setting them
% off from the rest of the text with a \clearpage as shown  below, and
% submit figures as separate files according to the Art Department's
% instructions.

\appendix
\section{Background}

% In this section, we first briefly discuss the dynamics of appraisals and risky decision making in teams. Then we provide the details of our experimental design. Finally, we describe the metrics for measuring model accuracy and how parameters are learnt.

%Finally, we present the four models for understanding decision making in mixed human-AI teams: the first two, {\bf NB} and {\bf CENT}, model the appraisal dynamics in teams and the last two, {\bf PT-NB} and {\bf PT-CENT}, model appraisal as well as risky decision-making.

\subsection{Appraisal Dynamics}

Individuals in a team appraise their own and others' displayed positions on issues. These appraisals may be based on individuals' task-relevant information, expertise, friendships, authority, or charisma~\cite{french1959bases}. A team's level task-performance depends on how these bases condition the allocation of relative influence to themselves and others ($0 \le w_{ij} \le 1$ $\forall i$, $\sum_{j=1}^{n}w_{ij} =1$ $\forall i$). The set of these weights defines a row-stochastic matrix $W$. This matrix is associated with an influence network $\mathcal{G}$ composed of $i \xrightarrow{w_{ij}>0} j$ arcs of $i$'s accorded relative direct influence to $j$. We assume that team's members' positions  evolve according to~\cite{degroot1974reaching} 

\[
X(k+1)=WX(k)=\ldots= W^{k}{X}(0), \;\; k=1,2, \dots ,
\]

\noindent where $X(0)$ represents the initial choices. If an initial consensus exists, then the prediction is that it will be maintained. If initial disagreement exists, then the existence of at least one globally reachable individual in $\mathcal{G}$ is a sufficient condition for the convergence of the influence system to consensus. A team with one globally reachable individual in $\mathcal{G}$ will form a consensus on the initial position of that individual. A team in which all individuals are mutually reachable in $\mathcal{G}$ will form a compromise consensus if $W$ is irreducible with at least one $0< w_{ii}< 1$. In all of the above cases, the convergence to consensus is associated with a  unique normalized left eigenvector of $W$ in which each eigenvalue is the total (direct and indirect) relative influence centrality of an individual's initial position on the team's consensus. Over a sequence of issues, the $W$ that is constructed by a team may differ from issue to issue, and along with such modifications the individuals' influence centralities may also vary over the  issue sequence.

%Individuals in a team appraise each other, leading to an influence network whose structure reflects the performance of individuals. Individuals also appraise the performance of agents.

%\textcolor{blue}{NEF:Option 1: I could write text on the French-Raven bases of social power (reward, punishment, authority, expert, referent and information power). In this case, these bases may be involved in determining relative weights. An eigenvector centrality measure can be obtained for the network of weights, but certain topological network conditions must be satisfied. This option seem to be what you are using.} 

%\textcolor{blue}{NEF:Option 2: I could write text that defines a mechanism of interpersonal influences and then derives individuals' influence centralities from the mechanism. Either the DeGroot model or its FJ model generalization might be used to do that. I prefer this option.} 

%\textcolor{blue}{NEF: If Option 1 is selected, then there needs to some data analysis that justifies the measure that you described. My concern is that your employed measure is weak and that its weakness is unduly elevating the value of the other models. 

\subsection{Prospect Theory}

In human decision making, the classic method of eliciting risk preference behavior is to present a sequence of choice dilemmas. A subject is presented with a series of problems and asked to select their preferred \textit{mixed gamble} in the set $\lbrace{{X = ((x_1,p_1),(x_2,p_2))}, {Y = ((y_1,q_1),(y_2,q_2))}\rbrace}$, where $x_1,y_1\geq 0$ are \$-valued gains and $x_2,y_2 < 0$ are \$-valued losses. Selecting gamble $X$ results in a payment of $\$x_1$ with probability $p_1$ or a debt of $\$x_2$ with probability $p_2$. Similarly for $Y$. Cumulative Prospect Theory proposes that individuals maintain an internal valuation function $V(X)$ assigning a psychological value to a set of risky outcomes called a prospect. The valuation function $V(X)$ for mixed gamble $X$ decomposes as  
\begin{equation}
V(X) = v^+(x_1)w^+(p_1) + v^-(x_2)w^-(p_2), \label{eq:cpt-1}
\end{equation}
where the value function $v^{+/-}(\cdot)$ and weighting function $w^{+/-}(\cdot)$ depend on the perception of the outcome relative to a reference point (i.e., is the outcome a gain or loss of capital, social status, etc.). 

Prospect theory defines a set of parameters  $\theta=(\alpha,\beta,\lambda,\gamma^+,\gamma^-)$, 
with the following natural interpretation~\cite{nilsson2011hierarchical}: $\alpha$ (resp. $\beta$) denotes sensitivity to gain (resp. loss)  outcomes, $\lambda$ denotes perceived impact of loss relative to gain,
$\gamma^+$ (resp. $\gamma^-$) denotes the degree to which gain (loss) probabilities are over- or under-weighted.
% Given two prospects $X$ and $Y$, the preference of an individual of $X$ over $Y$, defined $p(X,Y)$, is given by a sigmoid function of the difference in valuations $V(X) - V(Y)$ as:
% \vsb 
% \begin{align}
%     &p_\theta(X,Y) = {1}/\big( 1+  \exp(-\lbrace V(X|\theta) - V(Y|\theta) \rbrace )\big) \label{eq:Logistic}  
% \end{align}
% where
Thus, the value function $v^{+/-}(\cdot)$ and weighting function $w^{+/-}(\cdot)$ are defined as follows:

% \vsa \vsa 
\begin{equation}
\begin{aligned}
v^+(x) &= x^{\alpha}, \enspace\quad &&\alpha > 0,\\
v^-(x) &= -\lambda |x|^{\beta},\enspace &&\lambda, \beta>0,\\
w^+(p) &= \exp\!\big(-\big(\log \frac{1}{p}\big)^{\gamma^+}\big),\enspace &&\gamma^+ > 0,\\
w^-(p) &= \exp\!\big(-\big(\log \frac{1}{p}\big)^{\gamma^-}\big), \enspace &&\gamma^- > 0 .
\end{aligned}
\end{equation}

\section{Alternative Loss Functions}
The main body of the paper considered a metric $L^{(1)}$ for measuring the accuracy of models. Here, we consider two other possible metrics (or loss functions). The first is $L^{(2)}$, the symmetric counterpart to $L^{(1)}$.

\begin{equation}
L^{(2)} = \min_{p\ s.t.\ \forall k:\ p_k \leq p_i} -\sum_{i=1}^n p_i\log(q_i)
\label{eq:case2_human_entropy_loss}
\end{equation}

We obtain an analytical solution to the above optimization problem as follows. Let $q_j$ be a maximum value in $q$. The distribution $p$ that minimizes the above loss function can be computed as follows (Theorem~\ref{the:h_p_q} in the appendix).
\begin{itemize}
\item $p = q, \text{if}\ i = j$
\item \text{Otherwise,}\\
\hspace*{0.5in}$p_k = (1/n), k \in q_H$\\
\hspace*{0.62in} $= 0$, otherwise \\
where $q_{H}$ consists of those indices $k$ for which $q_k \geq q_i$, $n$ is the size of $q_{H}$.
%    \item $p_i=p_j$, $p_k=0, \forall k \neq i\neq j$, if $q_k \leq \sqrt{q_i q_j}$ .
\end{itemize}
%The value $\zeta$ is obtained by normalizing $p$ to 1. 

We also consider binary loss $L^{(b)}$:
\begin{equation}
L^{(b)} = -\sum_{i=1}^n p_i\log(q_i)
\label{eq:binary_loss}
\end{equation}
where $p_i=1$ if the team choose the $i$-th action, otherwise $p_i=0$.

It should be noted that the $L^{(2)}$ loss values for the {\bf PT-NB} and {\bf PT-CENT} models do not decrease in the same manner as that for $L^{(1)}$ loss values. This can be explained by considering how the second loss function is minimized. If the set $q_H$ (Eq.~\ref{eq:case2_human_entropy_loss}) does not change then the value of $L^{(2)}$ also does not change (we find that $q_H$ does not change between the non-PT and PT-based models in about half of the test cases). On the other hand, $L^{(1)}$ is more sensitive and it can change even if the set $q_H$ does not change. Tables~\ref{tab:loss_L2} and~\ref{tab:sl_1_L2} show the loss values corresponding to $L^{(b)}$ and $L^{(2)}$ for the four models and their significance comparisons in the first decision task, respectively. The orderings of the values in the two tables are roughly similar to those in Tables 1 and 2 presented earlier.
\begin{table}[H]
    \centering
    \begin{tabular}{|l|l|l|}
    \hline
        Model &Loss $L^{(b)}$ & Loss $L^{(2)}$\\ \hline
        {\bf NB} &$1.28 \pm 0.41$ &$1.22 \pm 0.21$\\
        {\bf CENT} &$1.17 \pm 0.34$ &$1.19 \pm 0.18$\\
        {\bf PT-NB} &$1.16 \pm 0.30$ &$1.10 \pm 0.29$\\
        {\bf PT-CENT} &$1.11 \pm 0.33$ &$1.00 \pm 0.31$\\ \hline
        {\bf RANDOM}  &$2.08 \pm 0$ &$2.08 \pm 0$\\ \hline
    \end{tabular}
    \caption{Loss values of the models in the first decision task.}
    \label{tab:loss_L2}
\end{table}

\begin{table}[H]
    \centering
    \begin{tabular}{|l|l|l|}
    \hline
        Wilcoxon signed-rank test& $L^{(b)}$ &$L^{(2)}$\\ \hline
        W({\bf NB},{\bf CENT}) &$<$0.01&0.12\\
        W({\bf NB},{\bf PT-NB}) &$<$0.01&$<$0.01\\
        W({\bf NB},{\bf PT-CENT}) &$<$0.01&$<$0.01\\
        W({\bf CENT},{\bf PT-NB}) &0.72&$<$0.01\\
        W({\bf CENT},{\bf PT-CENT}) &0.01&$<$0.01\\
        W({\bf PT-NB}, {\bf PT-CENT}) &0.17&$<$0.01\\\hline
    \end{tabular}
    \caption{Significance level ($p$-value) of the Wilcoxon signed-rank test for models in Decision Task 1. The null hypothesis is that the two related paired samples come from the same distribution.}
    \label{tab:sl_1_L2}
\end{table}

Table~\ref{tab:loss_task2_L2} shows the loss values corresponding to $L^{(b)}$ and $L^{(2)}$ for the models {\bf NB}, {\bf NB-H}, {\bf NB-A}, {\bf CENT}, {\bf CENT-H}, and {\bf CENT-A} in the second decision task. Table~\ref{tab:sl_2_L2} shows the corresponding significance comparisons. The orderings of values in the two tables again reflect that in Tables 3 and 4, suggesting robustness of the results.

\begin{table}[H]
    \centering
    \begin{tabular}{|l|l|l|}
    \hline
        Model &Loss $L^{(b)}$ &Loss $L^{(2)}$\\ \hline
        {\bf NB} &$0.50 \pm 0.27$ &$0.54 \pm 0.16$\\
        {\bf NB-H} &$1.43 \pm 0.37$ &$1.15 \pm 0.29$\\
        {\bf NB-A} &$0.64 \pm 0.47$ &$0.71 \pm 0.33$\\ \hline 
        {\bf CENT} &$0.59 \pm 0.46$ &$0.48 \pm 0.20$\\
        {\bf CENT-H} &$1.24 \pm 0.41$ &$1.09 \pm 0.26$\\
        {\bf CENT-A} &$0.67 \pm 0.39$ &$0.84 \pm 0.27$\\ \hline
        {\bf RANDOM} &$1.39 \pm 0$ &$1.39 \pm 0$\\\hline
    \end{tabular}
    \caption{Loss values of models in the second decision task.}
    \label{tab:loss_task2_L2}
\end{table}

\begin{table}[H]
    \centering
    \begin{tabular}{|l|l|l|}
    \hline
        Wilcoxon signed-rank test&$L^{(b)}$ &$L^{(2)}$\\ \hline
        W({\bf NB},{\bf NB-A}) &0.51&$<$0.01\\
        W({\bf NB},{\bf NB-H}) &$<$0.01&$<$0.01\\
        W({\bf NB-A},{\bf NB-H}) &0.02&$<$0.01\\\hline
        W({\bf CENT},{\bf CENT-A}) &0.51&$<$0.01\\
        W({\bf CENT},{\bf CENT-H}) &0.04&$<$0.01\\
        W({\bf CENT-A},{\bf CENT-H}) &0.06&$<$0.01\\\hline
    \end{tabular}
    \caption{Significance level ($p$-value) of the Wilcoxon signed-rank test for models in Decision Task 2. The null hypothesis is that the two related paired samples come from the same distribution.}
    \label{tab:sl_2_L2}
\end{table}

\section{Proofs of Lemmas and Theorems}
\begin{lemma}\label{lem:avg2nd}
For any $0 \leq x_1 \leq x_2 \leq 1$ and $0 \leq y_2 \leq y_1 \leq 1$, \[-(y_1 \log x_1 + y_2 \log x_2) \geq -(y_1 + y_2) \log \frac{x_1+x_2}{2} \]
\end{lemma}

\begin{proof}
Since the function $\log(x)$ is concave,

\begin{align*}
&\log \frac{x_1+x_2}{2} \geq \frac{1}{2}\log x_1 + \frac{1}{2}\log x_2\\
&(y_1 + y_2) \log \frac{x_1+x_2}{2} \geq \frac{y_1+y_2}{2}\log x_1 + \frac{y_1+y_2}{2}\log x_2\\
&= y_1 \log x_1 + y_2 \log x_2 + \frac{y_2 - y_1}{2}\log x_1 + \frac{y_1 - y_2}{2}\log x_2\\
&= y_1 \log x_1 + y_2 \log x_2 + \frac{y_2 - y_1}{2}(\log x_1 - \log x_2)\\
&= y_1 \log x_1 + y_2 \log x_2 + \frac{y_2 - y_1}{2}\log \frac{x_1}{x_2}\\
&\geq y_1 \log x_1 + y_2 \log x_2
\end{align*}
\end{proof}

The following lemma gives us an optimal solution to $\min_{q}H(p,q)$ over all distributions with mass $m_q$. 

\begin{lemma}
\label{lem:min_p_q}
Given a distribution $q$ with a total mass $m_q=\Sigma_i q_i$, among all distributions with a mass of $m_p$, the distribution that minimizes $-\Sigma_i q_i\log p_i$, is $p^* = \frac{m_p}{m_q}q$.
\end{lemma}

\begin{proof}
Consider any $p$ with $m_p=\Sigma_i p_i$. 
\begin{align*}
    - \Sigma_i p_i \log q_i &=     - \Sigma_i p_i \log \frac{q_i}{m_q} - \Sigma_i p_i \log m_q\\
    &= - m_p\Sigma_i \frac{p_i}{m_p} \log \frac{q_i}{m_q} - m_p \log m_q\\
    &= m_p H(\frac{p}{m_p}, \frac{q}{m_q}) - m_p \log m_q\\ 
    &\geq m_p H(\frac{p}{m_p}) - m_p \log m_q\\ 
\end{align*}

Since $p^* = \frac{m_p}{m_q}p$, it follows that
\begin{align*}
    - \Sigma_i p_i \log q^*_i &= m_p H(\frac{p}{m_p}, \frac{q^*}{m_q}) - m_p \log m_q\\ 
    &= m_p H(\frac{p}{m_p}, \frac{p}{m_p}) - m_p \log m_q\\    
    &= m_p H(\frac{p}{m_p}) - m_p \log m_q\\ 
\end{align*}
Therefore, 
$    - \Sigma_i p_i \log q_i \geq    - \Sigma_i p_i \log q^*_i$. 
 
\end{proof}

%The above lemma implies that given a probability distribution $p$ with a maximum value at $i$, that if we know an optimal solution $q$ has $q_j = \eta$, then the remaining entries $q_k = p_k \frac{1-\eta}{\sum_{r \neq j}p(r)}$. We also know that $q_j \geq q_i \geq q_k$ $k \neq i, k \neq j$, so $q$ satisfies the constraints when \[\frac{1-\eta}{1-p_j}p_i \leq \eta\]

The above two lemmas imply that given a probability distribution $q$, in order to construct a distribution $p$ whose maximum value is constrained to be at $i$ and whose cross-entropy $H(q,p)$ is minimized with respect to $q$, values in $p$ must be chosen such that $p_i = p_j = \eta$ if $q_i \leq q_j$, and the remaining mass is distributed over $p$ according to the shape of $q$. The following theorem establishes the optimal $p$ based on these considerations.

\begin{theorem}
Given a distribution $q$, a distribution $p$ with a maxima at $i$ that minimizes the cross-entropy $H(q,p)$ is defined as follows:\\
\hspace*{0.5in}$p_k = \eta$ if $k \in q_{H}$, \\
\hspace*{0.62in} $= q_k$, otherwise \\
where $q_{H}$ consists of those indices $k$ for which $q_k \geq q_i$, $n$ is the size of $q_{H}$, and $\eta = \frac{\sum_{k \in q_{H}}q_k}{n}$.
\label{the:h_q_p}
\end{theorem}

\begin{proof}
From Lemma~\ref{lem:avg2nd}, $p_i = p_k = \alpha, k \in q_H$ for optimal $H(q,p)$. Furthermore, from Lemma~\ref{lem:min_p_q}, for optimal $H(q,p)$, for any $k \not \in q_H$, $p_k = (1-n\alpha)\frac{q_k}{m}$, where $m = \Sigma_{l \not \in q_H} q_l$. Thus, at any optimal $p$,  

\begin{align*}
H(q,p) & = - (\sum_{k \in q_H} q_k) \log \alpha - \sum_{k \not \in q_H}q_k\log((1-n\alpha)\frac{q_k}{m})\\
\frac {dH(q,p)}{d\alpha}& = - \frac{\sum_{k \in q_H} q_k}{\alpha} + \frac{ n\sum_{k \not \in q_H}q_k}{1-n\alpha}\\
& =  - \frac{\sum_{k \in q_H} q_k}{\alpha} + \frac{nm}{1-n\alpha}\\
& =  \frac{n\alpha(m+\sum_{k \in q_H} q_k)-\sum_{k \in q_H} q_k}{\alpha(1-n\alpha)}\\
& =  \frac{n\alpha-\sum_{k \in q_H} q_k}{\alpha(1-n\alpha)}\\
\end{align*}
Note that at any feasible solution, $0 < \alpha \leq \frac{1}{n}$. The above derivative is negative when $\alpha < \eta$ and positive when $\alpha > \eta$. Furthermore, when $\alpha = \eta$, $p_k = q_k < \eta, \forall k \not \in q_H$. Therefore, $\alpha = \eta$ is a feasible solution. 
\end{proof}

\begin{lemma}
For any $0 \leq x_1 \leq x_2 \leq 1$ and $0 \leq y_2 \leq y_1 \leq 1$, \[-(x_1 \log y_1 + x_2 \log y_2) \geq -(\frac{x_1+x_2}{2} \log y_1 + \frac{x_1+x_2}{2} \log y_2) \]
\label{lemma1:v2}
\end{lemma}

\begin{proof}
\begin{align*}
&-(x_1 \log y_1 + x_2 \log y_2) + (\frac{x_1+x_2}{2} \log y_1 + \frac{x_1+x_2}{2} \log y_2) \\
&= \frac{x_2-x_1}{2} \log y_1 - \frac{x_2-x_1}{2} \log y_2\\
&= \frac{x_2-x_1}{2} (\log y_1 - \log y_2)\geq 0
\end{align*}
\end{proof}
%current
%The above lemma implies that a distribution $q$ constrained to $q_j \geq q_k,k\neq j$ and which is an optimal solution to $\min_{q}H(q,p)$, must satisfy $q_r \geq q_k$ if $p_r \geq p_k$. 
%reindex p in decreasing order
%The following theorem establishes the optimal $q$ based on these considerations.

\begin{theorem}
Given a distribution $q$, a distribution $p$ with a maxima at $i$ that minimizes the cross-entropy $H(p,q)$ is defined as follows:\\
\hspace*{0.5in}$p_k = (1/n)$ if $k \in q_{H}$, \\
\hspace*{0.62in} $= 0$, otherwise \\
where $q_{H}$ consists of those indices $k$ for which $q_k \geq q_i$, $n$ is the size of $q_{H}$.
\label{the:h_p_q}
\end{theorem}
\begin{proof}
Combining our constraints $p_i \geq p_k$ with Lemma~\ref{lemma1:v2} implies that at optimal $p$, $p_i = p_k, k \in q_H$. Next, we analyze which of the remaining entries in $p$ should be non-zero. Consider when a transfer of mass $\alpha$ from $\lbrace p_k \rbrace_{k \in q_H}$ to $p_l, l \notin q_H$,  reduces the entropy $H(p,q)$. For this, we must have\\
\begin{align*}
    - \alpha\log q_l&< - (\sum_{k \in q_H}(\alpha/n) \log q_k )\\
     -  \log q_l &< - \frac{1}{n} \sum_{k\in q_H}\log q_k
\end{align*}

Since $q_k > q_l, k \in q_H, l \notin q_H$, this statement is always false. Hence, all of the mass of $p$ will concentrate on $q_H$ and be uniformly distributed. Since there are $n$ elements of $q_H$, then $p_k = (1/n), \forall k \in q_H$ is an optimal solution.

\end{proof}

\end{document}